\documentclass[a, pdf]{iucr}              

\usepackage{color,xcolor}

\usepackage{xspace}
\RequirePackage{graphicx}
\usepackage{epstopdf}
\usepackage{mathtools, amsmath, amsthm, amssymb, amsfonts}
\usepackage[12pt]{moresize}
\usepackage{multirow}
\usepackage{enumitem}
\usepackage{url}

\newcommand{\be}{\begin{equation}}
\newcommand{\ee}{\end{equation}}

\newcounter{saveenumi}

\newcommand{\bes}{\begin{enumerate}[wide, labelwidth=!, labelindent=0pt, label=\textbf{\textcolor{blue}{\arabic*}.}]}
\newcommand{\ees}{\end{enumerate}}

\newcommand{\est}{$\sim$}

\makeatletter
\newcommand{\floatcaption}{%
\ifx \@captype \@undefined \@latex@error {\noexpand \caption outside float}\@ehd \expandafter \@gobble \else \refstepcounter \@captype \expandafter \@firstofone \fi {\@dblarg {\@caption \@captype }}%
}%
\makeatother

\definecolor{sobcolor}{HTML}{176d1b}

\definecolor{dgreen}{HTML}{008000}

\newcommand{\qmax}{\ensuremath{Q_{\mathrm{max}}}\xspace}
\newcommand{\qmin}{\ensuremath{Q_{\mathrm{min}}}\xspace}

\newcommand{\srmise}{\textsc{SrMise}\xspace}
\newcommand{\parscape}{\textsc{ParSCAPE}\xspace}

\journalcode{A}              

\begin{document}                  



\title{Algorithm for distance list extraction from pair distribution functions}
\shorttitle{Distance extraction from PDFs}


\author[a]{Ran}{Gu}{}{}
\author[a]{Soham}{Banerjee}{}{}
\cauthor[a]{Qiang}{Du}{qd2125@columbia.edu}{}
\author[a,b]{Simon J. L.}{Billinge}{}{}
\aff[a]{Department of Applied Physics and Applied Mathematics, Fu Foundation School of Engineering \& Applied Sciences,
Columbia University, \country{USA}}
\aff[b]{
    {Condensed Matter Physics and Materials Science Department,
    Brookhaven National Laboratory},
    \city{Upton, NY~11973}, \country{USA}}










\maketitle                        


\begin{abstract}
We present an algorithm to extract the distance list from atomic pair distribution functions (PDFs) in a highly automated way.
The algorithm is constructed via curve fitting based on a Debye scattering equation model. Due to the non-convex nature of the resulting optimization problem, a number of techniques are developed to overcome various computational difficulties.
A key ingredient is a new approach to obtain a reasonable initial guess based on the theoretical properties of the mathematical model.
Tests on various nanostructured samples show the effectiveness of the initial guess and the accuracy and overall good performance of the extraction algorithm.
This approach could be extended to any spectrum that is approximated as a sum of Gaussian functions.
\end{abstract}


\section{Introduction}

Determining the three-dimensional atomic positions in a nanostructure is one of the great challenges in materials science and engineering~\cite{billinge2007problem}. 
One experimentally accessible encoding of the local structure is the atomic pair distribution function (PDF), which is fundamentally a list of inter-atomic distaces in the material~\cite{egami;b;utbp12,warren2012x}.
This fact has led to a mathematical description of the nanostructure inverse problem as the unassigned distance geometry problem (uDGP)~\cite{billi;4or18,duxbu;dam16,duxbu;4or16,juhas;jac10,juhas;n06}. 
The PDF can be obtained by taking a Fourier transform of the structure function, which is extracted from the measured total scattering of x-rays, neutrons or electrons from a sample. 
The PDF method is widely used to study nanostructures~\cite{egami;b;utbp12,billinge2004beyond,billinge2007problem,young;jmc11,proff;jac97,page;jac11i,cliff;prl10}

The experimental peak width is determined by both physical properties and experimental resolution~\cite{egami;b;utbp12}.
In high symmetry structures such as bulk Ni, which crystallizes in a face-centered cubic lattice, distances of the same length occur frequently and the degeneracy of each distance can be estimated from the integrated area of each peak~\cite{egami;b;utbp12}.
However, a major challenge in determining the list of inter-atomic distances from a measured PDF comes from the fact that different interatomic vectors with similar lengths cannot be resolved due to peak overlap.
This is not a problem if we have a good structural model which can be fit to the data, which is the basis of PDF fitting programs such as PDFgui~\cite{billi;b;lsfd98,proff;jac99,farro;jpcm07}, an approach that is the real-space equivalent of Rietveld refinement of powder diffraction data~\cite{rietv;jac69}.
However, it presents a significant problem for programs that extract peak positions and intensities in the absence of a structural model, which would be the real-space equivalent of LeBail~\cite{lebail;mrb87} and Pawley~\cite{pawle;jac81} refinement in the powder diffraction world.

A program for extracting distance-lists from measured PDFs has been reported.
\parscape is an algorithm which can extract this complete set of information from the PDF by using the information-theoretic Akaike information criterion (AIC)~\cite{granlund2015algorithm}, available as a program \srmise on Diffpy.org.
However, the PDF baseline must be specified before peak extraction, and results are conditioned upon it.
The correct estimations of PDF baselines, especially from nanoparticle PDFs, remains challenging and requires human intervention, which is a drawback preventing full automation of \srmise.

Developing an algorithm for peak extraction which is automatable and robust to details of the baseline is our main goal here.
From the curve fitting point of view, an estimated distance list can be regarded as a variable to generate a simulated PDF based on the given mathematical model, then one may minimize the residual of the simulated PDF with respect to a target PDF to obtain the optimized distance list.
However the resulting curve fitting is generically a non-convex programming problem.
In order to solve the problem more effectively, we analyze the properties of the mathematical model which allows us to construct, automatically, the initial guess of the variables, with good fitting results demonstrated in preliminary tests using simulated and experimental PDF data.

This paper is organized as follows: In Section~\ref{sec:2}, we briefly introduce the PDF method and present the mathematical model that we use to approximate the experimental PDF. Section~\ref{sec:03} is a theory section containing the analysis of properties of the mathematical model.
This leads to an approach to guess the initial values of all variables. Section~\ref{sec:4} describes the formulation of the PDF distance list optimization. In Section~\ref{sec:5}, we present results from simulated and experimental PDF datasets used for testing the algorithm, and Section~\ref{sec:6} contains a summary of the main points of the paper.

\section{Mathematical Model of PDF}\label{sec:2}

We consider a nanostructure with a set of atoms. Let $N$ be the total number of atoms in the structure,
and $\{r_j\}_{j=1}^N$ denote  the positions of  the atoms.
The ideal PDF is defined by~\cite{farro;aca09,egami;b;utbp12}
\be \label{eq:idealpdf}
g(r)=\frac{1}{r}\frac{1}{N\langle f\rangle^2}\sum\limits_{j\neq l} f_j^* f_l \delta(r-r_{jl}).\ee
Here, for $j = 1\ldots N$ and $l = 1, \ldots N$,  $r_{jl}$ is the distance between atoms
$j$ and $l$ located at positions $r_j$, and $r_l$ so that
$r_{jl} = ||r_l - r_j ||$, where $||.||$ is the Euclidean norm,   $f_j$ is the scattering power of the atom at position $r_j$, and $f^\star_j$ is its complex conjugate.

The ideal PDF \eqref{eq:idealpdf} may also be obtained from measured data according to
\be G(r)=\frac{2}{\pi}\int_{0}^{\infty} F(Q) \sin (Qr) dQ, \ee
where $F(Q) = Q[S(Q)-1]$ is the normalized and corrected powder diffraction intensity, which is expressed in the Debye Scattering Equation as
\be\label{eq:idealf} F(Q) = \frac{1}{N\langle f\rangle^2}\sum_{l\neq j}f_j^\star f_l\frac{\sin(Qr_{jl})}{r_{jl}}.\ee
Here the quantity $S(Q)$ is called the structure function  and $F(Q)$ the reduced structure function~\cite{warren2012x} and
$Q$ is the magnitude of the scattering vector.

Due to physical constraints in the experiment, the variable $Q$ takes only values in the interval $[Q_{\min},Q_{\max}]$.
Thus, different from the standard Fourier transform, the PDF is obtained by the integral on this confined interval,~\cite{farro;aca09}
\be\label{eq:sinft}
G(r)=\frac{2}{\pi}\int_{Q_{\min}}^{Q_{\max}} F(Q) \sin (Qr) dQ.
\ee
To compute $G(r)$ numerically from $F(Q)$ on the discrete $Q_i$ grid, we approximate $G(r)$ by using the finite sum
\be\label{eq:sinftnum}
\ G(r)\approx\frac{2}{\pi}\sum\limits_{i=1}^{N_Q} F(Q_i) \sin (Q_ir) \Delta Q_i,
\ee
where $N_Q$ is the number of discrete values of $Q$, and $\Delta Q_i$ is the difference between two adjacent values of $Q$.

By ignoring the finite $Q_{\max}$, $G(r)$ and the radial distribution function $R(r)$ are related
by
\be
G(r)=\frac{R(r)}{r}-4\pi \rho_0 \gamma_0(r)r
\ee
where $\rho_0$ is the average density and $\gamma_0$ is the characteristic
function of the sample shape~\cite{fournet1955small,farro;aca09}. The term
$$4\pi \rho_0 \gamma_0(r)r=\frac{2}{\pi}\int_{0}^{Q_{\min}} F(Q) \sin (Qr) dQ$$
is a baseline, and we may think of $G(r)$ as the baseline plus peaks. In the literature to-date the shape of the baseline is either determined directly from the shape of the structural model~\cite{juhas;aca15,farro;jpcm07}, or approximated using expansions of ad hoc mathematical functions~\cite{korsunskiy2005exact,neder2005structure,korsunskiy2007aspects,neder2007structural}. In the case of bulk crystals, $\gamma_0(r)\approx 1$, is a linear baseline~\cite{egami1998local,proffen1999pdffit,farro;jpcm07}.  However in general, without a good structural model, the baseline is not known \textit{a priori}.

The experimental signal is a time and ensemble average of large numbers of atoms and the Dirac delta-function peaks given in the ideal PDF \eqref{eq:idealpdf} broaden into nearly Gaussian peaks.
In reciprocal space, to account for atomic motion, \eqref{eq:idealf} is replaced
by a version that includes Debye-Waller effects,
\be\label{eqn:fqmultielem}
F(Q) = \frac{1}{N\langle f\rangle^2}\sum_{l\neq j}f_j^\star f_l(e^{-\frac{1}{2}\sigma_{jl}^2Q^2})\frac{\sin(Qr_{jl})}{r_{jl}}.
\ee
Here, $\sigma_{jl}$ is the correlated broadening factor for the atom pair~\cite{proffen1999pdffit,thorpe2002semiconductors,jeong2003lattice}. This mathematical model has been successfully used
to study nanostructures by a number of authors~\cite{zhang2003water,cervellino2006efficient}.

For the case of samples made of a single atom type, the atomic form factors $f_j$ can be factored out resulting in new functions
\be\label{eqn:g1}
\hat F(Q)=\frac{F(Q)}{f_j^\star f_l} \quad \mbox{ and }\quad
\hat G(r)=\frac{2}{\pi}\int_{Q_{\min}}^{Q_{\max}} \hat F(Q) \sin (Qr) dQ.
\ee
We regard
\be
\label{eqn:f1}
\hat F(Q)=\frac{1}{N\langle f\rangle^2}\sum_{l\neq j}(e^{-\frac{1}{2}\sigma_{jl}^2Q^2})\frac{\sin(Qr_{jl})}{r_{jl}}
\ee
as our mathematical model. If the material contains different atomic types, we use Equation~\eqref{eqn:fqmultielem} instead.

In Equation~\eqref{eqn:f1}, we want to merge distances of the same length together. Notice that distances of the same length may have different $\sigma$. Nevertheless, we still put them together because peaks at the same position are more difficult to differentiate. We then obtain the following mathematical model,
\be\label{eqn:f2}
\hat F(Q)=\sum\limits_{i=1}^k \frac{m_i}{r_i}e^{-\frac{1}{2}\sigma^2_{i}Q^2}\sin(Qr_{i}).
\ee
Here, $k$ is the number of different values of unresolved distances. $m_i$ represents the relative multiplicity which is equal to multiplicity times $1/(N\langle f\rangle^2)$ .

During curve fitting, we first determine the value of $k$. Then we recognize all the $r_i$, $m_i$, $\sigma_i$ as variables. In the next section, we discuss how to construct the initial guesses for these variables.

\section{Mathematical Model Analysis and Initial Guess}\label{sec:03}

This section is divided into several parts. We first present some properties of the mathematical model used to calculate PDFs, and then describe a few approaches for determining an initial guess distance list, using different atomic structures as examples.

\subsection{Properties}\label{sec:prop}

\begin{figure}
\includegraphics[width=1\textwidth]{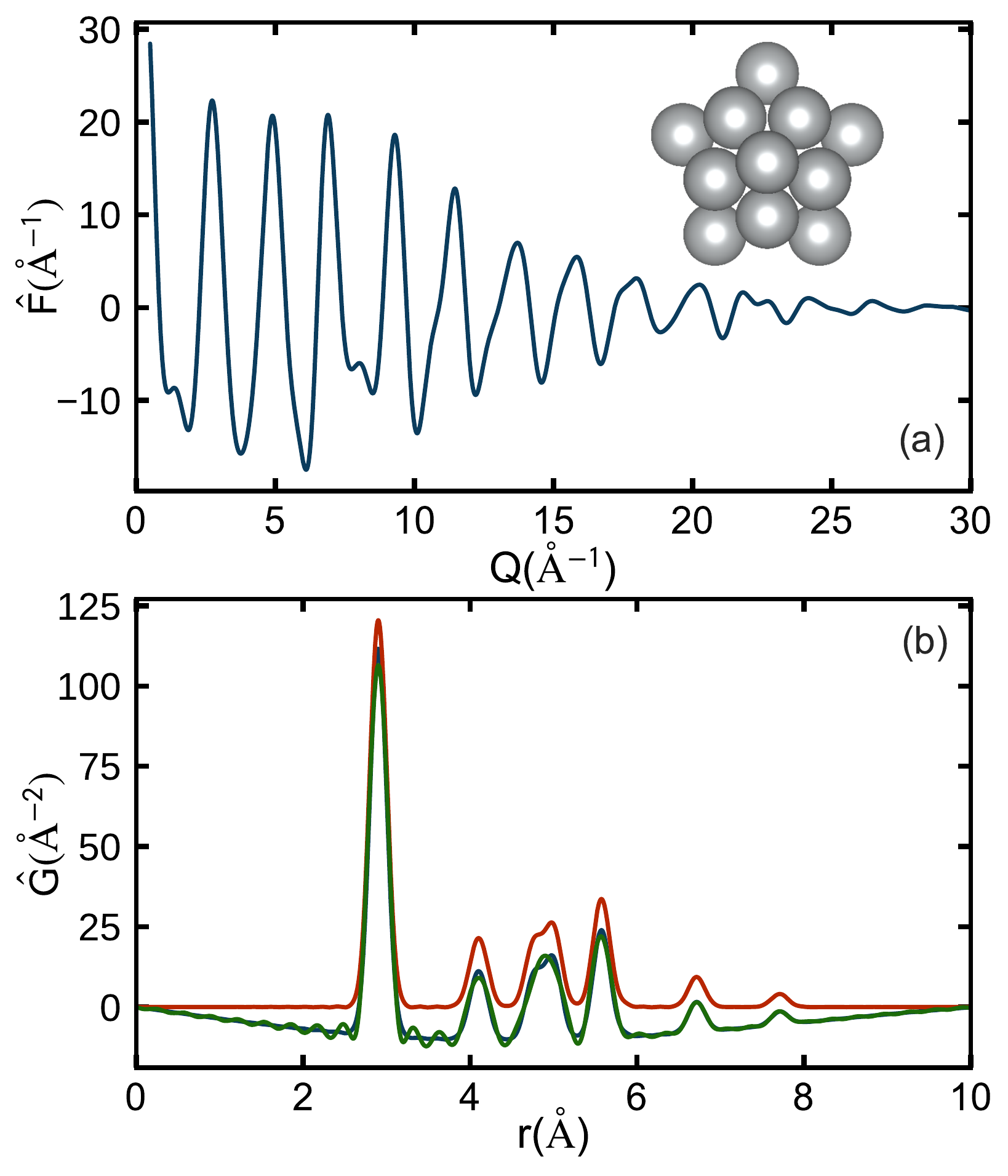}
\caption{(a) Simulated $F(Q)$ for an 18-atom Lennard-Jones decahedron (shown in inset) using Equation~\eqref{eqn:f2} with fixed $\sigma_i=0.1~\text{\AA}$. (b) PDFs after Fourier transformation of the simulated $F(Q)$ in (a) following Equation~\eqref{eq:sinftnum} and using ($Q_{\min}$, $Q_{\max}$) values, in units~\AA$^{-1}$: red (0, 30), dark blue (0.5, 30), and green (0.5, 20).}\label{fig:pdfqminqmaxeffect}
\end{figure}

In real experiments, the intensities are measured only over a range $Q_{min}<Q<Q_{max}$, which introduces aberrations to the data that must be handled by our automated algorithm.
To explore this in more detail we first consider a low energy 18-atom Lennard-Jones decahedral cluster~\cite{wales2001cambridge}.
Figure~\ref{fig:pdfqminqmaxeffect}(a) shows the function $\hat F(Q)$, calculated from Equation~\eqref{eqn:f2} using the decahedral structure model, over the $Q$-range from $Q_{\min}=0~\text{\AA}^{-1}$ to $Q_{max}=30~\text{\AA}^{-1}$,  with $\sigma_i$ set to $0.1$~\AA.
The three curves in the bottom panel of Figure~\ref{fig:pdfqminqmaxeffect} show the function $\hat G(r)$, calculated from Equation~\eqref{eq:sinftnum} where $\sigma_i$ is kept fixed at $0.1$~\AA, and the $Q$-ranges are varied in order to illustrate the effects on the transformed PDFs.
There are a number of peaks in $\hat G(r)$ where each peak represents one or more distances.
We take as reference a PDF calculated with a small but finite \qmin, and a large \qmax ($Q_{\min}=0.5~\text{\AA}^{-1}$, $Q_{\max}=30~\text{\AA}^{-1}$) which is shown as the dark blue curve, and compare it with the red curve ($Q_{\min}=0~\text{\AA}^{-1}$, $Q_{\max}=30~\text{\AA}^{-1}$), where we find that a larger \qmin makes the $\hat G(r)$ baseline deeper, as expected~\cite{farro;aca09,egami;b;utbp12}.
Comparing the same dark blue curve with the green curve ($Q_{\min}=0.5~\text{\AA}^{-1}$, $Q_{\max}=20~\text{\AA}^{-1}$), we find that smaller \qmax leads to larger oscillations in $\hat G(r)$, again as expected~\cite{egami;b;utbp12}.
The coordinates of the atoms in the structure models were determined algorithmically using the Atomic Simulation Environment (ASE) Python package~\cite{Larsenatomicsimulationenvironment2017}, as described in~\cite{BanerjeeImprovedModelsMetallic2018a}.

To consider the effects of the $Q$-range we decompose the PDF into different contributions from the different ranges of $Q$.
According to Equations~\eqref{eqn:g1} and~\eqref{eqn:f2}, we have
\begin{align}
\hat G(r)&=
\frac{2}{\pi}\int_{Q_{\min}}^{Q_{max}}\sum\limits_{i=1}^k \frac{m_i}{r_i}e^{-\frac{1}{2}\sigma^2_{i}Q^2}\sin(Qr_{i})\sin(Qr)dQ\\
&=\frac{2}{\pi} \left\{ \int_{0}^{\infty}-\int_{0}^{Q_{\min}}-\int_{Q_{max}}^{\infty}\right\} \sum\limits_{i=1}^k \frac{m_i}{r_i}e^{-\frac{1}{2}\sigma^2_{i}Q^2}\sin(Qr_{i})\nonumber\\
&\hskip 36pt\sin(Qr)dQ\\
&= \hat G_1(r) + \hat G_2(r) + \hat G_3(r)
\end{align}
Here we split $\hat G(r)$ to three parts. If $Q_{max}$ is large enough, we regard the integral from ${Q_{max}}$ to $\infty$ as 0 due to the exponential term. The termination of the Fourier transform varies with the type of material and with the amplitude of lattice vibrations but, in general, termination with $Q > 30~\text{\AA}^{-1}$ produces minimal errors~\cite{toby1992accuracy}. However, for lower ${Q_{max}}$ the ripples may be signicant (e.g., green PDF in the bottom panel of Figure~\ref{fig:pdfqminqmaxeffect}), in which case, in the absence of a structural model, the oscillations may be mistaken as physical peaks. This increases the computational effort for extraction. We can use the following inequality as a threshold to reduce the number of mis-identified peaks,
\begin{align}
 \left|\hat G_3(r)\right| =&\left|\frac{2}{\pi} \int_{Q_{max}}^{\infty}\sum\limits_{i=1}^k\frac{m_i}{r_i}e^{-\frac{1}{2}\sigma^2_{i}Q^2}\sin(Qr_{i})\sin(Qr)dQ\right| \nonumber \\
 \leq & \frac{2}{\pi}\sum\limits_{i=1}^k \frac{m_i}{r_i}\int_{Q_{max}}^{\infty}e^{-\frac{1}{2}\sigma^2_{i}Q^2}dQ\\
 \leq & \frac{1}{\pi}\frac{N}{r_1\langle f\rangle^2}\int_{Q_{max}}^{\infty}e^{-\frac{1}{2}\underline{\sigma}^2Q^2}dQ
 \label{eq:g3}
\end{align}
where $r_1$ is the smallest distance and $\underline{\sigma}$ is a lower bound of all $\sigma_i$.

The unattenuated term contains the atomic-scale structural information and is given by
\begin{align}
\hat G_1(r)&=\frac{2}{\pi}\int_{0}^{\infty}\sum\limits_{i=1}^k \frac{m_i}{r_i}e^{-\frac{1}{2}\sigma^2_{i}Q^2}\sin(Qr_{i})\sin(Qr)dQ\\
\label{eq:sumGauss}&=\frac{1}{\sqrt{2\pi}}\sum\limits_{i=1}^k \frac{m_i}{r_i\sigma_i}\left(e^{-\frac{(r-r_i)^2}{2\sigma_i^2}}-e^{-\frac{(r+r_i)^2}{2\sigma_i^2}}\right),
\end{align}
which is a sum of Gaussians as expected.

\begin{figure}
\includegraphics[width=1\textwidth]{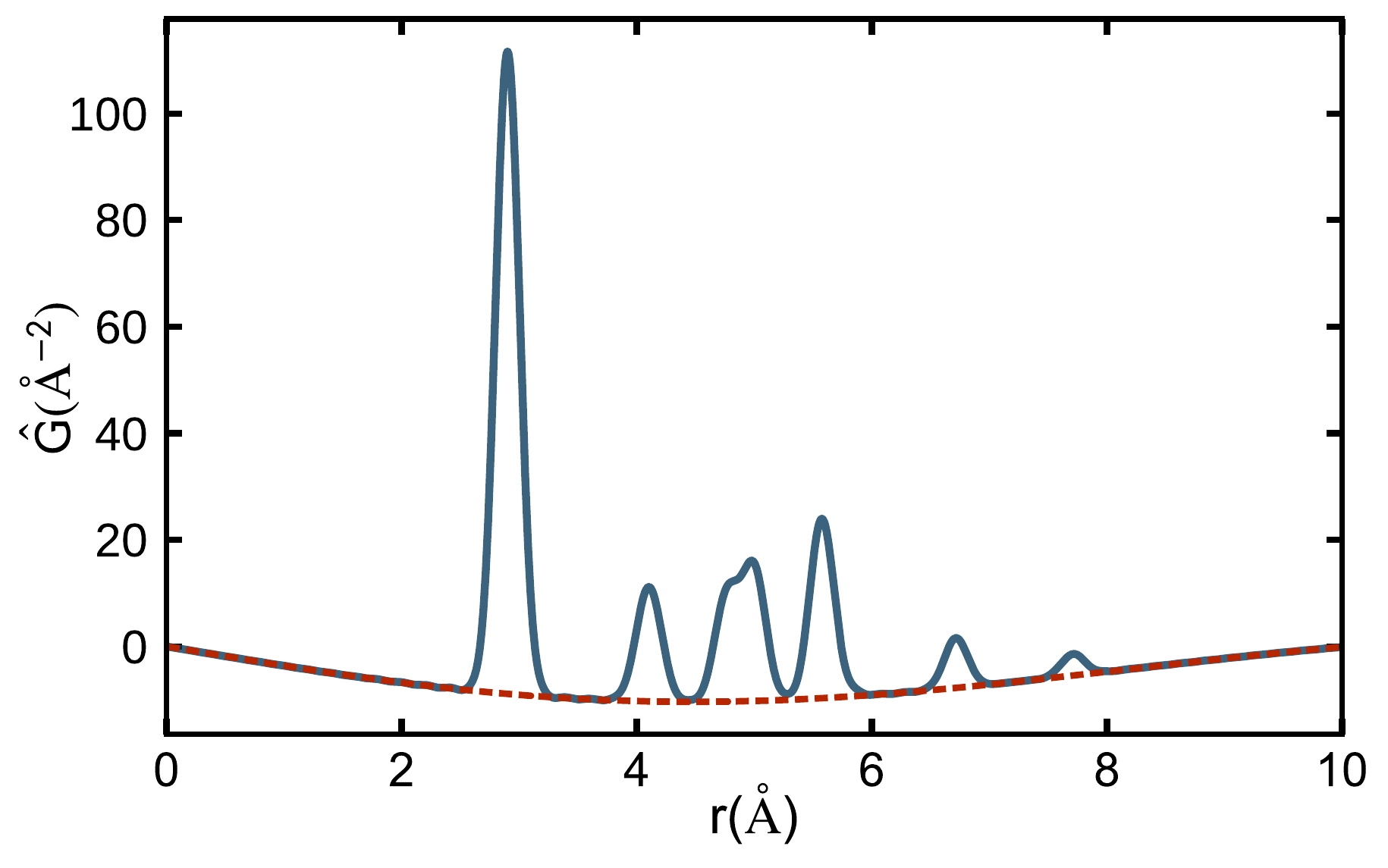}
\caption{The blue curve is the simulated PDF for an 18-atom Lennard-Jones decahedron transformed from the $F(Q)$ shown in Figure~\ref{fig:pdfqminqmaxeffect}, with the following configuration: $Q_{\min}=0.5~\text{\AA}^{-1}$, $Q_{\max}=30~\text{\AA}^{-1}$, $\sigma_i=0.1~\text{\AA}$. The red dashed line is an approximation to the baseline for $\hat G(r)$ following Equation~\eqref{eq:bline}, using the same values used for the simulated PDF in blue.}\label{fig:pdfbaseline}
\end{figure}

The part that determines the baseline is given by~\cite{farro;aca09}
\begin{align}
\hat G_2(r)=& \frac{2}{\pi}\int_{0}^{Q_{\min}}\sum\limits_{i=1}^k \frac{m_i}{r_i}e^{-\frac{1}{2}\sigma^2_{i}Q^2}\sin(Qr_{i})\sin(Qr)dQ\\
=& \frac{2}{\pi}\int_{0}^{Q_{\min}}\sum\limits_{i=1}^k \frac{m_i}{r_i}(1+O(\sigma^2_{i}Q^2))\sin(Qr_{i})\sin(Qr)dQ\\
=& \frac{1}{\pi}\sum\limits_{i=1}^k \frac{m_i}{r_i}\left(\frac{\sin((r-r_{i})Q_{\min})}{r-r_{i}}-\frac{\sin((r+r_{i})Q_{\min})}{r+r_{i}}\right)\\
& +\sum\limits_{i=1}^k \frac{m_i\sigma_i^2}{r_i}O(Q_{\min}^3)\,\nonumber
\end{align}
which has been simplified here by taking terms only up to second order in a Taylor series expansion. Approximating the baseline by
\be \label{eq:bline} \frac{1}{\pi}\sum\limits_{i=1}^k \frac{m_i}{r_i}\left(\frac{\sin((r-r_{i})Q_{\min})}{r-r_{i}}-\frac{\sin((r+r_{i})Q_{\min})}{r+r_{i}}\right)\ee
does provide a good approximation as shown in Figure~\ref{fig:pdfbaseline}.

\subsection{Initial Guess on a single peak}\label{sec:singlepeak}

If the target function is a sum of Gaussians, one can make an initial guess using the second order derivative~\cite{goshtasby1994curve}.
Approaches using higher order derivatives and wavelet transforms have been applied to a variety of experimentally measured spectra to find local maxima, often combined with gaussian denoising filters~\cite{HuangPrecisionPeakDetermination1988,Gregoirewavelettransformalgorithm2011,SavitzkySmoothingDifferentiationData1964}.
However, our mathematical model is not exactly a sum of Gaussians.  The Gaussians are contained in $\hat{G}_1(r)$, but in general our signal also includes a baseline term, $\hat G_2(r)$, and termination effects, $\hat{G}_3(r)$.  We now consider the effects from these different $Q$-dependent contributions on our ability to accurately extract peak parameters using second and higher order derivatives.

We begin by considering $\hat G_1(r)$, which contains the structural signal.
The second derivative of $\hat G_1(r)$, following Equation~\eqref{eq:sumGauss}, is
\begin{align}
\begin{split}
\hat G_1''(r)=\frac{1}{\sqrt{2\pi}}\sum\limits_{i=1}^k \frac{-m_i}{r_i\sigma_i^3}\left[\left(1-\frac{(r-r_i)^2}{\sigma_i^2}\right)e^{-\frac{(r-r_i)^2}{2\sigma_i^2}} \right. \\
\left. -\left(1-\frac{(r+r_i)^2}{\sigma_i^2}\right)e^{-\frac{(r+r_i)^2}{2\sigma_i^2}}\right].
\end{split}
\end{align}
First, we take the simplest case of an isolated single peak that we label as the $i$-th peak.
Due to the characteristics  of the exponential function, other peaks are some distance away and may affect the single peak slightly. Then,
$r_i$ can be extracted by using the location of the local maximum of $\hat G(r)$ or $-\hat G''(r)$. Consider $r$ around $r_i$,
\begin{align}
\hat G_1''(r)=\frac{1}{\sqrt{2\pi}} \frac{-m_i}{r_i\sigma_i^3}\left(1-\frac{(r-r_i)^2}{\sigma_i^2}\right)e^{-\frac{(r-r_i)^2}{2\sigma_i^2}}.
\end{align}
We have two zero crossing points of $\hat G_1''(r)$, $z_1^\star=r_i-\sigma_i$ and $z_2^\star=r_i+\sigma_i$. Then we take as the initial guess of $\sigma_i$,
\be\sigma_i=\frac{z_2^\star-z_1^\star}{2}.\label{eqn:g1zcross}\ee
This result will be accurate if the curved baseline contribution to $\hat G$,  $\hat G_2(r)$, does not introduce a significant shift on the zero crossing points.

Consider the second order derivative of $\hat G_2(r)$,
\begin{align}
\hat G_2''(r)=&  \frac{2}{\pi}\int_{0}^{Q_{\min}}-Q^2\sum\limits_{i=1}^k \frac{m_i}{r_i}e^{-\frac{1}{2}\sigma^2_{i}Q^2}\sin(Qr_{i})\sin(Qr)dQ\\
=& \frac{2}{\pi}\int_{0}^{Q_{\min}}-Q^2\sum\limits_{i=1}^k \frac{m_i}{r_i}(1+O(\sigma^2_{i}Q^2))\sin(Qr_{i})\sin(Qr)dQ
\end{align}
{\small
\begin{align}
=& \frac{1}{\pi}\sum\limits_{i=1}^k \frac{m_i}{r_i}
\left[ \left(\frac{-Q_{\min}^2\sin((r-r_i)Q_{\min})}{r-r_i}+\frac{-2Q_{\min}\cos((r-r_i)Q_{\min})}{(r-r_i)^2}\right.\right.\\
&\left.+\frac{2\sin((r-r_i)Q_{\min})}{(r-r_i)^3}\right) -\left(\frac{-Q_{\min}^2\sin((r+r_i)Q_{\min})}{r+r_i}  \right.\nonumber \\
&\left.\left. +\frac{-2Q_{\min}\cos((r+r_i)Q_{\min})}{(r+r_i)^2}+\frac{2\sin((r+r_i)Q_{\min})}{(r+r_i)^3}\right)\right]+O(Q_{\min}^5) \nonumber
\end{align}
}
where we have again used the  Taylor expansion.
Define $\eta=(r-r_i)Q_{\min}$.  If $\eta$ is close to zero, by Taylor expansion,
\begin{align}
&\frac{-Q_{\min}^2\sin((r-r_i)Q_{\min})}{r-r_i}+\frac{-2Q_{\min}\cos((r-r_i)Q_{\min})}{(r-r_i)^2}\\
&+\frac{2\sin((r-r_i)Q_{\min})}{(r-r_i)^3} \nonumber \\
 = &-Q_{\min}^3\frac{\sin(\eta)}{\eta}  - 2Q_{\min}^3 \frac{\cos(\eta)\eta -\sin(\eta)}{\eta^3}\\
=&-\frac{1}{3}Q_{\min}^3+O(\eta^2Q_{\min}^3).
\end{align}
On the other hand,  
if $\eta$ is away from zero, then
\begin{align}
\frac{-Q_{\min}^2\sin((r-r_i)Q_{\min})}{r-r_i}&+\frac{-2Q_{\min}\cos((r-r_i)Q_{\min})}{(r-r_i)^2}\\
+\frac{2\sin((r-r_i)Q_{\min})}{(r-r_i)^3} &= O(Q_{\min}^3).\nonumber
\end{align}
Therefore,
\begin{equation}\label{eq:g2esti}
\hat G_2''(r)=O(Q_{\min}^3).
\end{equation}
This is small and we can, with confidence, set $z^\star$ and $z$ to be the zero crossings on the same side of $\hat G_1''(r)$ and $\hat G_1''(r)+\hat G_2''(r)$, respectively.  Further, we can ignore $G_3''$ because when $Q_{max}$ is large enough, we can approximate it as zero due to its exponential term. This means that although $G(r)$ is not purely a sum of Gaussians, the multiple derivative zero crossings method can still be expected to give acceptably good initial estimates of single peak positions. Then, we have
\be\hat G_1''(z^\star)=0,\qquad \hat G''(z)=0.\ee
and by the Mean Value Theorem,
\be z-z^\star=\frac{\hat G_1''(z)-\hat G_1''(z^\star)}{\hat G_1'''(\hat z)}=\frac{-\hat G_2''(z)}{\hat G_1'''(\hat z)},\ee
where $\hat z$ is a real number between $z$ and $z^\star$. For $r$ around $r_i$,
\be \hat G_1'''(r)=\frac{1}{\sqrt{2\pi}}\frac{m_i}{r_i\sigma_i^4}\left(3\frac{r-r_i}{\sigma_i}-\frac{(r-r_i)^3}{\sigma_i^3}\right)e^{-\frac{(r-r_i)^2}{2\sigma_i^2}}.\ee
When $r\in(r_i-\sqrt{2}\sigma_i,r_i-0.2\sigma_i)\cup(r_i+0.2\sigma_i,r_i+\sqrt{2}\sigma_i)$,
\be |\hat G_1'''(r)|>\frac{1}{\sqrt{\pi}}\frac{m_i}{r_i\sigma_i^4}e^{-1}\; .\ee
Therefore, we have
\be z-z^\star=O(\sigma_i^4 Q_{\min}^3),\label{eqn:g2zcross}\ee
\begin{equation}\label{eq:zcross}
\sigma_i=\frac{z_2-z_1}{2}+O(\sigma_i^4 Q_{\min}^3).
\end{equation}

To compute different even order derivatives of $\hat G(r)$ numerically, we do not use finite difference approximations, because it is very easy to produce numerical instabilities when calculating higher order derivatives. Instead, we calculate the derivatives directly on the formula
\be
\hat G(r)=\frac{2}{\pi}\int_{Q_{\min}}^{Q_{\max}} \hat F(Q) \sin (Qr) dQ.
\ee
We only consider even order derivatives of $\hat G(r)$, which give the peaked functions that we seek for the zero crossing analysis.  Intuitively, $r^\star$ is the maximizer/minimizer of $\hat G^{(2s)}(r)$ and then $\hat G^{(2s)+1}(r^\star)=0$. Therefore,
\be
\hat G^{(2s)}(r)=\frac{2}{\pi}\int_{Q_{\min}}^{Q_{\max}} (-1)^sQ^{2s}\hat F(Q) \sin (Qr) dQ.
\ee
When the measured $\hat F(Q)$ is taken only at a finite point $Q_i$, we approximate
\be \label{eq:g2s}
\hat G^{(2s)}(r)\approx\frac{2}{\pi}\sum\limits_{i=1}^{N_Q} (-1)^sQ_i^{2s}\hat F(Q_i) \sin (Q_ir) \Delta Q_i,
\ee
where $N_Q$ is the number of discrete values of $Q$, and $\Delta Q_i$ is the difference between two adjacent Q. The high order derivatives obtained from $\hat F(Q)$ by this method are more stable. When $\hat F(Q)$ is not available from experimental data, a finite difference approximation can be used to calculate higher order derivatives from experimentally measured $G(r)$.

\begin{figure}
\includegraphics[width=1\textwidth]{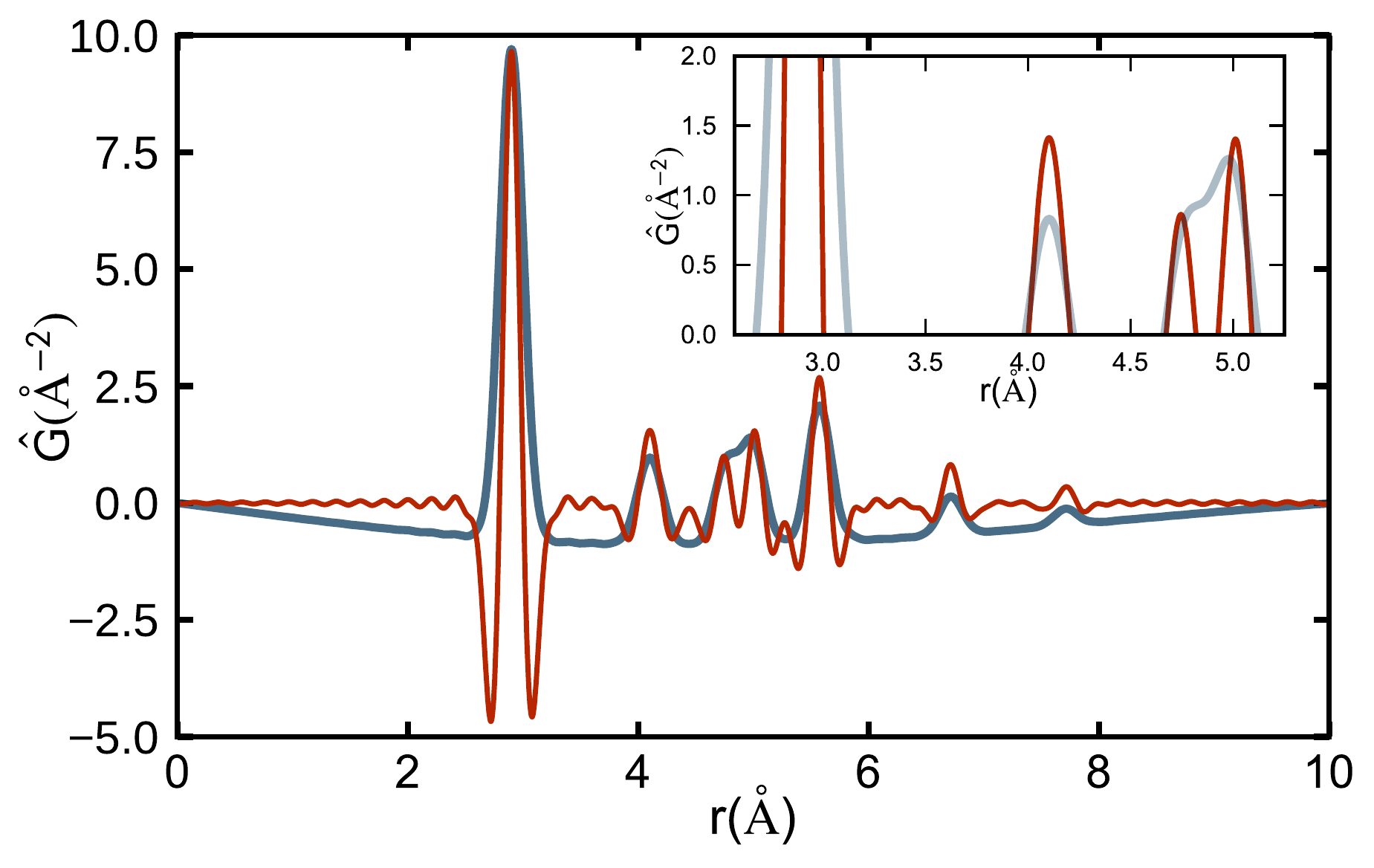}
\caption{The blue curve is the simulated PDF for an 18-atom Lennard-Jones decahedron as shown in Figures~\ref{fig:pdfqminqmaxeffect}-\ref{fig:pdfbaseline}. Overlaid in red is the scaled second order derivative, $-\hat G''(r)$, calculated by Equation~\eqref{eq:g2s}. The inset is a zoom-in to the zero crossings before \est5.3~\AA.}\label{fig:pdf2ndderivative}
\end{figure}

In Figure~\ref{fig:pdf2ndderivative}, the blue curve, $\hat G(r)$,  is the simulated PDF of an 18-atom decahedron introduced in Section~\ref{sec:prop}, with $Q_{\min}= 0.5~\text{\AA}^{-1}$, $Q_{\max}= 30~\text{\AA}^{-1}$ and $\sigma_i=0.1~\text{\AA}$ for all $i=1,\ldots,k$.
The red curve, $-\hat G''(r)$, is calculated as given in Equation~\eqref{eq:g2s}. A magnified view is shown in the inset from \est2.5-5.3~\AA. The distances between the zero crossing of the first two peaks are very close to $0.2~\text{\AA}$ which is twice the value of $\sigma$ used to generate this PDF. This shows that it is reasonable to guess the initial value of $\sigma_i$ using the zero-crossing estimation given in Equation~\eqref{eqn:g1zcross}.

\subsection{Initial guess for the case of overlapped peaks}

Figure~\ref{fig:pdf2ndderivative} shows another interesting phenomenon. The overlapped peaks near $4.8$~\AA\ appear as two well separated peaks in the second derivative curve shown in red. This motivates us to use higher derivatives to search for  overlapped peaks.

We consider the $n$-th derivative, where $n$ is an even number, $n=2s$.
Using the fact that
\be
(e^{-\frac{x^2}{2}})^{(n)}=(-1)^s\sum\limits_{j=0}^s (-1)^j C_n^{2j}(n-2j-1)!!x^{2j}e^{-\frac{x^2}{2}},\ee
where $C_n^k$ is the number of combinations of $n$ items taken $k$ at a time which is defined as $n!/k!(n-k)!$, $!$ is factorial, and $!!$ is double factorial. Considering $r$ around $r_i$ and following Equation~\eqref{eq:sumGauss},
we again take a Taylor expansion giving
\begin{align}
\hat G_1^{(n)}(r)=&(-1)^s\frac{1}{\sqrt{2\pi}} \frac{m_i}{r_i\sigma_i^{n+1}}(n-1)!!\\
&(1-s\frac{(r-r_i)^2}{\sigma_i^2}+O(\frac{(r-r_i)^4}{\sigma_i^4}))e^{-\frac{(r-r_i)^2}{2\sigma_i^2}} \nonumber.
\end{align}
The two nearest zero crossing points of $\hat G_1^{(n)}(r)$ are $z_1^\star\approx r_i-\frac{1}{\sqrt{s}}\sigma_i$ and $z_2^\star\approx r_i+\frac{1}{\sqrt{s}}\sigma_i$. Similar to Equation~\eqref{eq:g2esti}, we obtain that $\hat G_2^{(n)}(r)$ does not affect this guess too much due to
\be\hat G_2^{(n)}(r)=O(Q_{\min}^{n+1}),\ee
we can again use
\be\sigma_i\approx \frac{z_2-z_1}{2}\sqrt{\frac{n}{2}}\ee
as the initial guess of $\sigma_i$.

\begin{figure}
\includegraphics[width=1\textwidth]{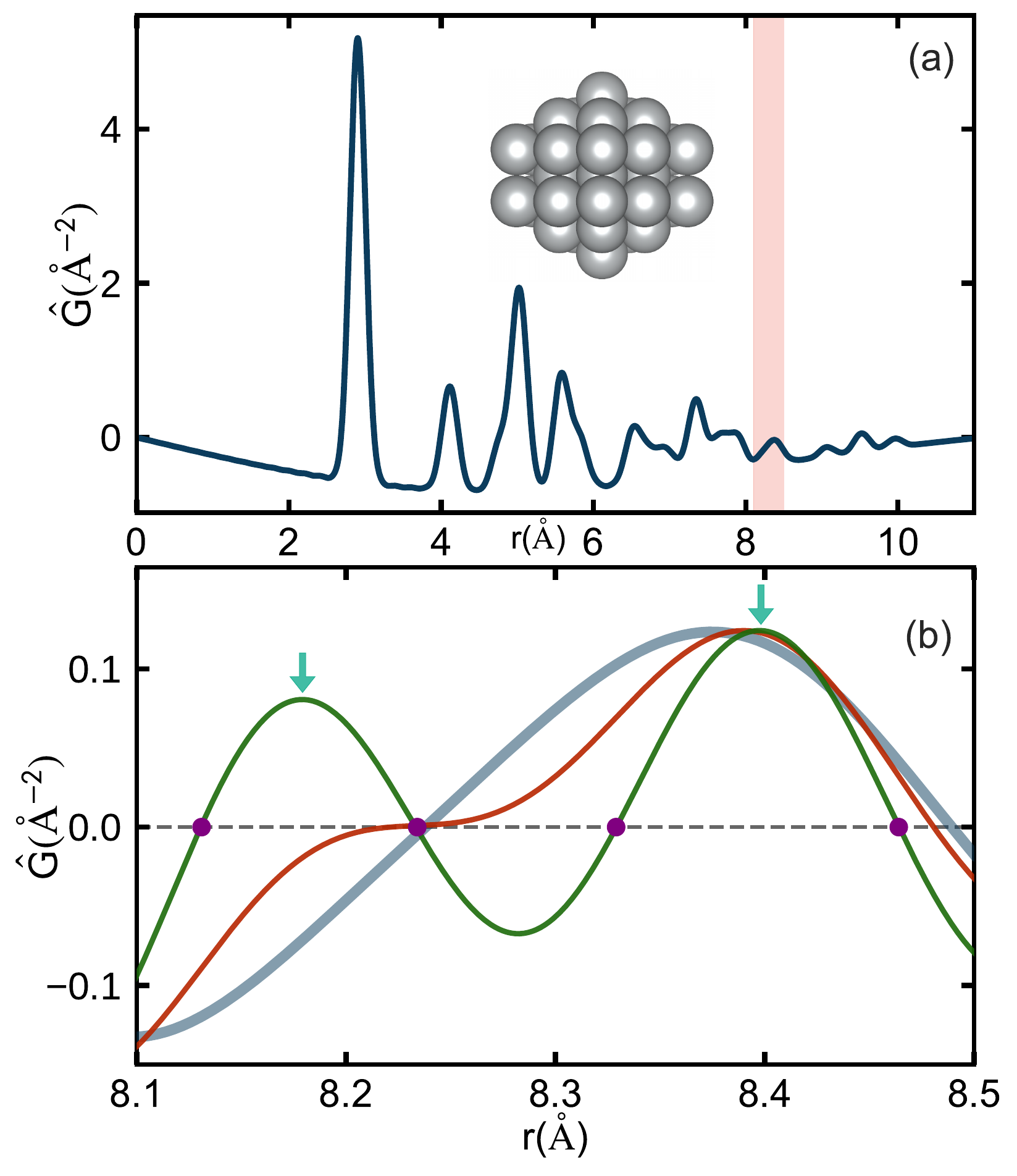}
\caption{(a) Simulated PDF (blue) from  a 39-atom decahedron, shown in the inset. (b): A magnified view of the highlighted $r$-range from (a) where the same simulated PDF is shown in light blue and overlaid with the 2nd (red) and 4th (green) order derivatives,  $-\hat G''(r)$ and $\hat G^{(4)}(r)$ respectively. Zero-crossings for $\hat G^{(4)}(r)$ are marked in purple and the initial guesses for the two overlapped peaks within this $r$-range are shown with teal arrows.}
\label{fig:pdfhighderivative}
\end{figure}

The PDF simulated from a different model, a 39-atom decahedral cluster, provides an illustration of another interesting point of using zero-crossings from higher order derivatives to locate peaks.
Figure~\ref{fig:pdfhighderivative}(a) shows the PDF calculated from this model.
We set all $\sigma_i$ to be $0.1~\text{\AA}$, $Q_{\min}=0.5~\text{\AA}^{-1}$, $Q_{\max}=30~\text{\AA}^{-1}$.

The peak in the interval from 8.1~\AA\ to 8.5~\AA\ looks single valued but contains two true peaks at 8.23~\AA\ and 8.40~\AA, respectively.
In Figure~\ref{fig:pdfhighderivative}(b) we magnify this narrow $r$-range and overlay the 2nd order (red) and 4th order (green) derivatives on top of the simulated PDF (light blue). This shows that the 2nd order derivative with only two zero-crossings cannot sufficiently resolve the split peak, whereas the 4th order derivative can.
In Figure~\ref{fig:pdfhighderivative}(b), the initial guesses for the position of the two peaks are highlighted with teal arrows, which are determined from the four zero-crossings (purple markers).

In this case, $\hat G^{(4)}(r)$, was optimal for separating these two peaks.
In practice, higher derivatives give greater selectivity for finding overlapped peaks, but also dramatically increases the number of zero crossings originating from noise, and a balance must be struck between these two competing factors.
As a rule of thumb, we have found that the $(n+2)$-th derivative should be considered only when the $n$-th derivative does not result in reasonable initial guesses for $r_i$ and $\sigma_i$.  In practice, for a fully automated peak extraction program we do not want human involvement in the decision making.
We have found that $\hat G^{(4)}(r)$ is a good balance between sensitivity and noise suppression in the examples we have tried.
In the future, we may experiment with different protocols, for example, adaptively trying derivatives of different order, and even changing the order used to extract signals from specific peaks in the PDF.
These improvements to the automated heuristic have not proven to be necessary to date.

\subsection{Initial guess for the peak amplitude, $m_i$}

After giving the initial guesses for $r_i$ and $\sigma_i$, we can estimate values for the peak amplitudes, $m_i$, by using the following standard box constrained least square,
\begin{align}
\min\limits_{m_1,\ldots,m_k}&
\displaystyle
\sum\limits_{j=1}^{n_Q}\left[\hat F(Q_j)-\sum\limits_{i=1}^k \frac{m_i}{r_i}e^{-\frac{1}{2}\sigma^2_{i}Q_j^2}\sin(Q_jr_{i})\right]^2\label{eqn:m1}
\\
s.t.\quad & m_i\geq 0,\ i=1,\ldots,k
\end{align}
where $n_Q$ is the number of discrete values of $Q$ and the $r_i$ and $\sigma_i$ values are held constant. This is a convex quadratic programming problem which can be easily solved without having to specify initial values for the $m_i$.  For example, here we use the primal-dual interior-point algorithm~\cite{nesterov1997self,nesterov1998primal}.

\section{Optimization}\label{sec:4}

With initial values for the variables we can continue to the optimization step.  A standard box constrained least square problem is used to fit either the $\hat F(Q)$ or the $\hat G(r)$ curve according to
\begin{align}
\label{eqn:fitFQ}
\min\limits_{r_i,\sigma_i,m_i}
\sum\limits_{j=1}^{nq}&\left[\hat F(Q_j)-\sum\limits_{i=1}^k \frac{m_i}{r_i}e^{-\frac{1}{2}\sigma^2_{i}Q_j^2}\sin(Q_jr_{i})\right]^2\\
s.t.\quad & m_i\geq 0,\ i=1,\ldots,k\nonumber
\end{align}
and
{\small
\begin{align}
\label{eqn:fit_Gr}
\min\limits_{r_i,\sigma_i,m_i}
\sum\limits_{l=1}^{nr}&\left[\frac{2}{\pi}\sum\limits_{j=1}^{nq} \left(\sum\limits_{i=1}^k \frac{m_i}{r_i}e^{-\frac{1}{2}\sigma^2_{i}Q_j^2}\sin(Q_jr_{i})\right)\sin(Q_j r_l)\Delta Q_j-G(r_l)\right]^2.\\
s.t.\quad & m_i\geq 0,\ i=1,\ldots,k\nonumber
\end{align}
}
We have found that compared to fitting on $\hat F(Q)$, the real space optimization is more computationally intensive, but can yield better solutions.
For the optimization we use a subspace trust-region method based on the interior-reflective Newton method described in~\cite{coleman1996interior}. Each iteration involves the approximate solution of a large linear system using the method of preconditioned conjugate gradients (PCG). Due to the high nonlinearity and non-convexity of this least square problem, the solutions calculated by the solver depend sensitively on the starting values.  Nonetheless, the initial values we obtained from the differential zero crossings have proven to be stable.

\section{Testing the approach}\label{sec:5}

Two target PDFs are tested for our extraction algorithm in this section. One shows the peaks extracted from a simulated PDF, and the other from an experimental PDF of atomically precise clusters where the structure has been satisfactorily solved.

\subsection{Test on simulated data.}

\begin{figure}
\includegraphics[width=1\textwidth]{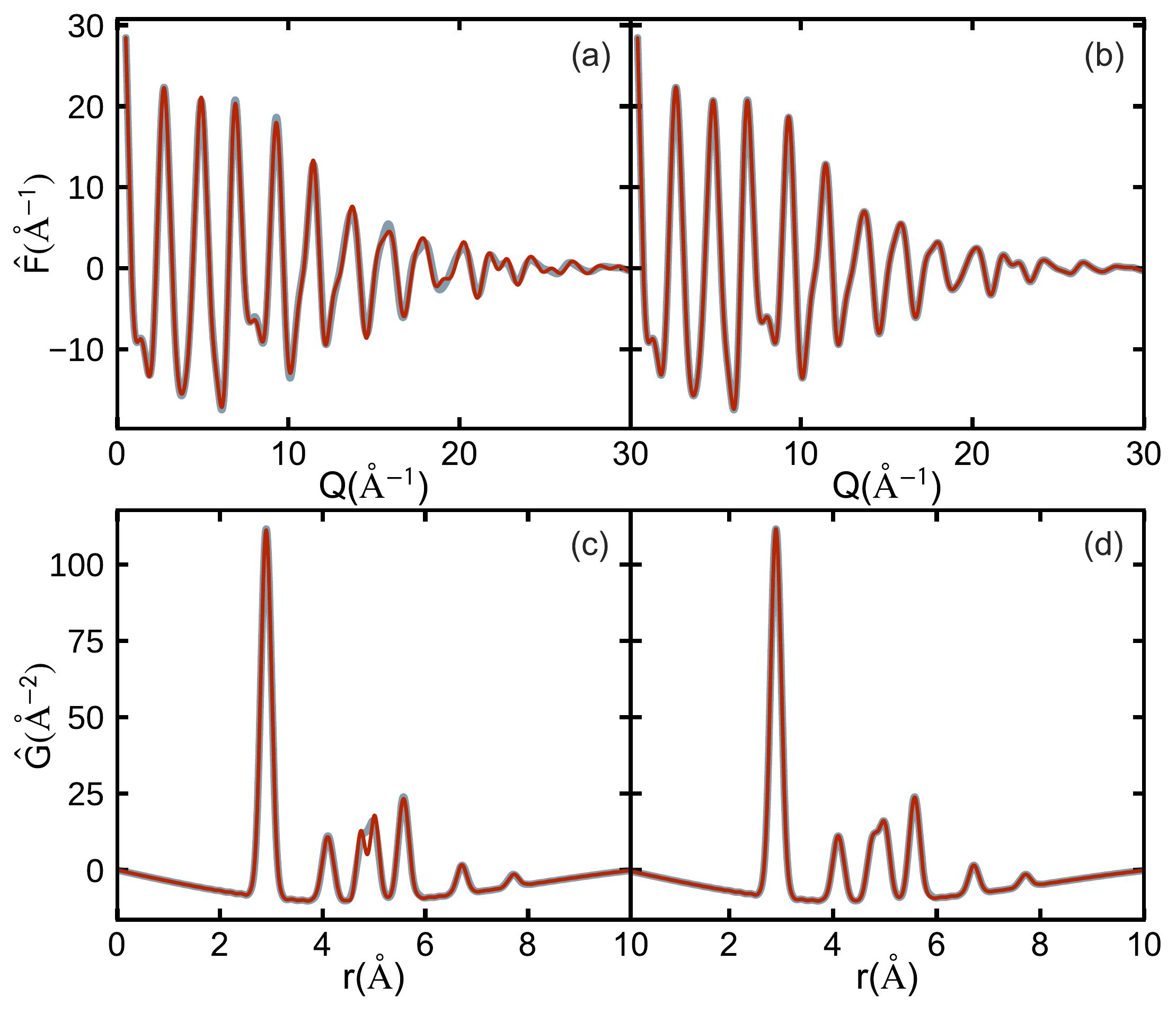}
\caption{Simulated (light blue) and extracted (red) $\hat F(Q)$ (top) and $\hat G(r)$ (bottom) curves from initial guesses (a,c) and following optimization (b, d) for an 18-atom Lennard-Jones decahedron.}
\label{fig:fitsimulationqmax30}
\end{figure}

We revisit the 18-atom Lennard-Jones decahedron discussed in Sections~\ref{sec:prop}-\ref{sec:singlepeak} to generate a simulated PDF for this test. Here we set $Q_{\min}=0.5~\text{\AA}^{-1}$, $Q_{\max}=30~\text{\AA}^{-1}$, and all $\sigma_i=0.1~\text{\AA}$.
The distances extracted using our approach are reproduced in Table~\ref{tab:merge} and the resulting PDF curves after the initial guess and the full refinement steps are shown in Figure~\ref{fig:fitsimulationqmax30}.
For this relatively high resolution ($Q_{max}=30$~\AA$^{-1}$) case the extraction is working rather well.  The only peaks that could not be extracted separately by the program were very close to each other and the extraction returns single peaks with the full integrated intensities of both unresolved distances.
The first three columns in the table show the ground-truth parameters we set to generate the PDF. There are 11 different distances, of which we find seven.
The program could not resolve the peaks at 2.8921~\AA\ and 2.9443~\AA\ in the first feature.
A single peak was returned at 2.9008~\AA, very close to the weighted average position of the ground-truth peaks, 2.9013~\AA. The program also returned a multiplicity of 56.9680, very close to the sum of the true multiplicities of the unresolved peaks, 57.  The second and fourth peaks were also unresolved doublets.  The widest unresolved splitting was 0.075~\AA.  The peak at 4.7640~\AA\ and the unresolved doublet at 4.99~\AA\ were successfully resolved and they are a little over 0.2~\AA\ apart, very close to the expected resolution of data with a $Q_{max}=30~\text{\AA}^{-1}$~\cite{farro;prb11}.

In general, we may be working with data that were measured at lower real-space resolutions, for example, $Q_{\max} = 23~\text{\AA}^{-1}$, which we also tested.
Smaller $Q_{\max}$ results in less resolution but also in termination ripples, or oscillations, that might confuse the extraction process.
In the initial guess stage, our program finds 36 peaks.
In fact, most of them are from termination ripples.
However, after the optimization stage, a subset of the peak amplitudes are close to 0.
In Table~\ref{tab:merge}, we filter these distances programmatically, and only list the peaks which return a multiplicity larger than 1 and compare them with the extracted peaks when $Q_{\max}=30~\text{\AA}^{-1}$.
The quality of the extraction is worse for $Q_{\max}=23~\text{\AA}^{-1}$ compared to $Q_{\max}=30~\text{\AA}^{-1}$, but it is still quite reasonable, containing a small number of false positives with very little weight.
The program still successfully resolved the peaks at 4.76~\AA\ and 4.99~\AA\ but this time it misassigned some weight between these two peaks.  However, all in all, it was a satisfactory extraction.

\paragraph{An example with experimental data.}

\begin{figure}
\includegraphics[width=1\columnwidth]{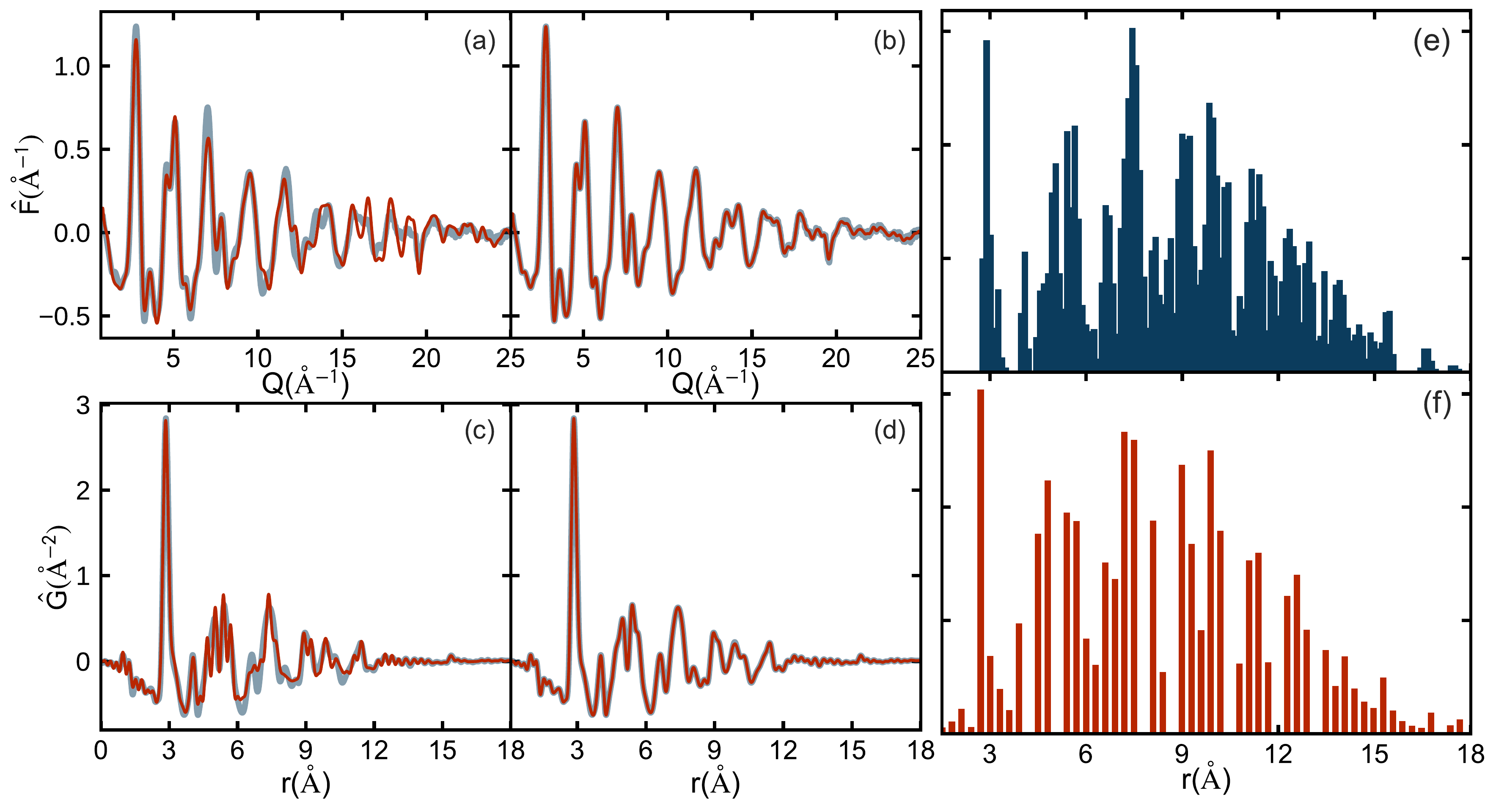}
\caption{
Experimental (light blue) and extracted (red) $\hat F(Q)$ (top) and $\hat G(r)$ (bottom) curves from initial guesses (a,c) and following optimization (b, d) for a Au$_{144}$(SC6)$_{60}$ cluster sample. Histograms from the~\cite{Lopez-AcevedoStructureBondingUbiquitous2009} structure solution distances (e) and the optimized-extracted distances (f) obtained from the fit to the experimental data shown in (d).}\label{fig:fitexperimentAu144}
\end{figure}

Next we test our distance extraction algorithm on experimental PDF data collected from 144-atom gold nanoclusters capped with 60 thiolate staples, Au$_{144}$(SC6)$_{60}$.
Sample preparation is described in~\cite{QianAmbientSynthesisAu1442011} and data acquisition and processing to obtain the PDF are described in~\cite{jense;nc16}.
A well-established DFT structure model exists for this sample~\cite{Lopez-AcevedoStructureBondingUbiquitous2009} (LA model) which was previously shown to be in good agreement with the measured PDFs~\cite{jense;nc16,BanerjeeImprovedModelsMetallic2018a}.
The relaxed LA cluster structure is complex, with low symmetry chiral and staple arrangements of shell atoms on top of a higher symmetry Mackay icosahedral core.
In Figure~\ref{fig:fitexperimentAu144} we show the PDF resulting from the initial guess (a, c) and final fitting  (b, d) to extract distances from the experimental $\hat F(Q)$ and $\hat G(r)$ curves. A $Q$-range from  $Q_{\min}=0.8~\text{\AA}^{-1}$ to $Q_{\max}=25~\text{\AA}^{-1}$ was used for the PDF transformation, and a 4th order derivative was used to obtain initial guesses for $r_i$ and $\sigma_i$.
\onecolumn
\twocolumn
To improve the extraction, we provided an additional constraint such that $\{\sigma_i\leq 0.16~\text{\AA}\}$ to bound all the $\sigma_i$'s.
Values above this bound would yield isotropic atomic displacement parameters (ADPs) greater than \est0.025~\AA$^2$, which are unphysically large for homogeneous, atomically precise nanocluster samples.
The fitted PDF and $F(Q)$ of the peak models converges nicely to the data, as evident in Figure~\ref{fig:fitexperimentAu144}.
The fit of the extracted PDF to the measured one indicates we have obtained good convergence, but it is not a measure of the quality of the extraction, which is rather determined by how well the extracted distances agree with the actual ones.  In Figure~\ref{fig:fitexperimentAu144}(e) and (f) we show histograms of the actual, and optimized-extracted distances, respectively from the Au$_{144}$(SC6)$_{60}$ experimental data.
A visual comparison suggests that the distribution of distances from the experimental data is very similar to the true distance histogram obtained from the LA structure solution; the overall shape and many of the fine features match well, albeit more coarse-grained for the extracted peaks due to unresolved overlapping peaks.
A truncated list ($2.68<r<7.81$~\AA) of the extracted peak parameters ($r_i$, $m_i$) is provided in Table~\ref{tab:au144vs}, in addition to the optimized $\sigma_i$ values needed to calculate $\hat F(Q)$ and $\hat G(r)$.
Due to the dense distribution of true distances, we assigned each extracted distance to a group of true distances, and summed the multiplicities per bin as the total multiplicity.
To determine estimated multiplicities from the experimental Au$_{144}$(SC6)$_{60}$ data, we performed an additional extraction from data simulated from the LA structure model. The simulated PDF was generated with $Q_{\min}=0.8~\text{\AA}^{-1}$, $Q_{\max}=25~\text{\AA}^{-1}$ and all $\sigma_i=0.1~\text{\AA}$. This normalization results in good agreement between the experimental and simulated multiplicities from the LA model. The results from the simulated extraction are also provided in Table~\ref{tab:au144vs} next to the experimental extraction.  This is a challenging low-symmetry nanostructure, but the auto extraction nonetheless seems to be working well.

\section{Conclusion}\label{sec:6}

In this paper, we have proposed an algorithm to extract distance lists from a target PDF with no {\it a priori} structural information. We use a mathematical model utilizing the sum of Guassians nature of the Debye scattering equation and the PDF to automatically recover peak position, and therefore interatomic distance information.
It firstly uses an automated approach to find an initial guess for all the variables and then solves a global optimization problem. The preliminary tests show the effectiveness of the initial guess and good performance and accuracy of the extraction.  The approach has been successfully tested on PDFs simulated from known nanoparticle clusters as well as from a challenging low-symmetry experimental dataset.




\ack{\paragraph{Acknowledgment}
The authors thank Chia-Hao Liu and Yunzhe Tao for useful discussions.
S.B and S.J.B. also thank Christopher J. Ackerson and Kirsten Marie Jensen for the synthesis and characterization of cluster samples.
This research is supported by the U.S. National Science Foundation (NSF) through grant DMREF-1534910.
S.B. acknowledges support from the National Defense Science and Engineering Graduate Fellowship (DOD-NDSEG) program.
Data collected at the Advanced Photon Source at Argonne National Laboratory was supported by the U.S. Department of Energy, Office of Science, Office of Basic Energy Sciences (DOE-BES), under contract number DE-AC02-06CH11357.}



\bibliographystyle{iucr}
\bibliography{bib_merged_final}

\begin{table*}
\centering
\floatcaption{True and extracted peak parameters for the simulated PDF of an 18-atom Lennard-Jones decahedron.}\label{tab:merge}
\begin{tabular}{|r|c|r|r|r|r|r|r|r|r|r|r|}
\hline
 \multicolumn{3}{|c|}{\multirow{2}[0]{*}{True values}} & \multicolumn{6}{c|}{$Q_{\max}=$30~\AA$^{-1}$} & \multicolumn{3}{c|}{$Q_{\max}=$23~\AA$^{-1}$} \\
 \cline{4-12}
 \multicolumn{3}{|c|}{}  & \multicolumn{3}{c|}{Initial guess} & \multicolumn{3}{c|}{Final extraction} & \multicolumn{3}{c|}{Final extraction}\\
 \hline
 \multicolumn{1}{|c|}{$r(\text{\AA})$} & \multicolumn{1}{c|}{$\sigma(\text{\AA})$} & \multicolumn{1}{c|}{$m$}& \multicolumn{1}{c|}{$r(\text{\AA})$}& \multicolumn{1}{c|}{$\sigma(\text{\AA})$} & \multicolumn{1}{c|}{$m$} & \multicolumn{1}{c|}{$r(\text{\AA})$}& \multicolumn{1}{c|}{$\sigma(\text{\AA})$} & \multicolumn{1}{c|}{$m$}& \multicolumn{1}{c|}{$r(\text{\AA})$}& \multicolumn{1}{c|}{$\sigma(\text{\AA})$} & \multicolumn{1}{c|}{$m$} \\
 \hline
 2.8921  & 0.1  & 47  & \multirow{2}[0]{*}{2.9000} & \multirow{2}[0]{*}{0.1040} & \multirow{2}[0]{*}{57.5527} & \multirow{2}[0]{*}{2.9008} & \multirow{2}[0]{*}{0.1019} & \multirow{2}[0]{*}{56.9680} & \multirow{2}[0]{*}{2.9007} & \multirow{2}[0]{*}{0.1023} & \multirow{2}[0]{*}{57.3677} \\
 2.9443  & 0.1  & 10  &       &       &       &       &       &       &       &       &  \\
 \hline
 4.0525  & 0.1  & 5  & \multirow{2}[0]{*}{4.1020} & \multirow{2}[0]{*}{0.1100} & \multirow{2}[0]{*}{15.2690} & \multirow{2}[0]{*}{4.1027} & \multirow{2}[0]{*}{0.1061} & \multirow{2}[0]{*}{14.9729} & \multirow{2}[0]{*}{4.1026} & \multirow{2}[0]{*}{0.1038} & \multirow{2}[0]{*}{14.7513} \\
 4.1271  & 0.1  & 10  &       &       &       &       &       &       &       &       &  \\
 \hline
       &       &       &       &       &       &       &       &       & 4.4167  & 0.2925  & 1.7502  \\
 \hline
 4.7640  & 0.1  & 15  & 4.7450  & 0.0800  & 14.1855  & 4.7647  & 0.1001  & 15.0521  & 4.7454  & 0.0891  & 11.9304  \\
 \hline
 4.9787  & 0.1  & 10  & \multirow{2}[0]{*}{5.0110} & \multirow{2}[0]{*}{0.0880} & \multirow{2}[0]{*}{19.8334} & \multirow{2}[0]{*}{4.9942} & \multirow{2}[0]{*}{0.1007} & \multirow{2}[0]{*}{19.8774} & \multirow{2}[0]{*}{4.9786} & \multirow{2}[0]{*}{0.1128} & \multirow{2}[0]{*}{23.6725} \\
 5.0092  & 0.1  & 10  &       &       &       &       &       &       &       &       &  \\
 \hline
 5.5732  & 0.1  & 30  & 5.5740  & 0.1045  & 31.0025  & 5.5755  & 0.1023  & 30.6557  & 5.5748  & 0.1036  & 31.4847  \\
 \hline
 5.7841  & 0.1  & 1  &       &       &       &       &       &       &       &       &  \\
 \hline
       &       &       &       &       &       &       &       &       & 6.2230  & 0.2271  & 1.4548  \\
 \hline
 6.7147  & 0.1  & 10  & 6.7120  & 0.0970  & 9.8322  & 6.7147  & 0.0996  & 9.9450  & 6.7168  & 0.1050  & 10.9121  \\
 \hline
       &       &       &       &       &       &       &       &       & 7.2787  & 0.1356  & 1.0148  \\
 \hline
 7.7084  & 0.1  & 5  & 7.7190  & 0.1000  & 4.9749  & 7.7084  & 0.0993  & 4.9490  & 7.7079  & 0.1093  & 5.9554  \\
 \hline
       &       &       &       &       &       &       &       &       & 8.6454  & 0.1834  & 1.2553  \\
 \hline
\end{tabular}%
\end{table*}

\begin{table*}
\centering
\floatcaption{Interatomic distances from the Lopez-Acevedo (LA) structure model (ground truth) and optimized-extracted peak parameters for the case of data simulated from the LA model, and from experimental PDF data from Au$_{144}$(SC6)$_{60}$ clusters}\label{tab:au144vs}
\begin{tabular}{|r|r|r|r|r|r|r|r|r|r|}
    \hline
	\multicolumn{2}{|c|}{LA model distances}&\multicolumn{4}{c|}{Simulated extraction}&\multicolumn{4}{c|}{Experimental extraction}\\
\hline
	\multicolumn{1}{|c|}{$r(\text{\AA})$}& \multicolumn{1}{c|}{$m$} & \multicolumn{1}{c|}{$r(\text{\AA})$}& \multicolumn{1}{c|}{$\sigma(\text{\AA})$}&\multicolumn{2}{c|}{$m$}&\multicolumn{1}{c|}{$r(\text{\AA})$}& \multicolumn{1}{c|}{$\sigma(\text{\AA})$}&\multicolumn{2}{c|}{$m$}\\
\hline
 2.68-3.06 & 528.00  & 2.87  &  0.13   & \multicolumn{2}{c|}{552.80} & 2.85  & 0.12  & \multicolumn{2}{c|}{565.68} \\
 \hline
 3.08-3.33 & 102.00  & 3.21   & 0.11  & \multicolumn{2}{c|}{105.76} & 3.15  & 0.14  & \multicolumn{2}{c|}{128.01} \\
 \hline
 3.35-3.49 & 12.00  & 3.45  &  0.09  & \multicolumn{2}{c|}{\ \ 16.95} & 3.40  & 0.13  & \multicolumn{2}{c|}{\ \ 73.61} \\
 \hline
       &       & 3.67  & 0.08  & \multicolumn{2}{c|}{\ \ \ \ 7.66} &       &       & \multicolumn{2}{c|}{} \\
       \hline
 3.82  & 1.00  &       &       & \multicolumn{2}{c|}{} & 3.79  & 0.12  & \multicolumn{2}{c|}{\ \ 38.66} \\
 \hline
 3.87-4.20 & 179.00  & 4.03   & 0.13  & \multicolumn{2}{c|}{198.15} & 4.03  & 0.11  & \multicolumn{2}{c|}{181.19} \\
 \hline
 4.30  & 1.00  &       &       & \multicolumn{2}{c|}{} &       &       & \multicolumn{2}{c|}{} \\
 \hline
 4.40-4.47 & 40.00  & 4.44  & 0.12  & \multicolumn{2}{c|}{\ \ 28.49} & 4.41  & 0.09  & \multicolumn{2}{c|}{\ \ 58.26} \\
 \hline
 4.49-4.80 & 214.00  & 4.64  & 0.16 & \multicolumn{2}{c|}{220.81} & 4.68  & 0.14  & \multicolumn{2}{c|}{270.63} \\
 \hline
 4.82-5.25 & 468.00  & 4.98  & 0.16  & \multicolumn{2}{c|}{497.69} & 5.00  & 0.14  & \multicolumn{2}{c|}{415.99} \\
 \hline
 5.26-5.54 & 387.00  &  5.41 &  0.13 & \multicolumn{2}{c|}{399.73} & 5.38  & 0.11  & \multicolumn{2}{c|}{363.49} \\
 \hline
 5.55-5.81 & 369.00  & 5.66  & 0.13  & \multicolumn{2}{c|}{355.96} & 5.64  & 0.13  & \multicolumn{2}{c|}{349.33} \\
 \hline
 5.83-6.11 & 105.00  & 5.93  & 0.16  & \multicolumn{2}{c|}{132.43} & 5.91  & 0.15  & \multicolumn{2}{c|}{126.89} \\
 \hline
 6.13-6.25 & 43.00  & 6.19  & 0.09  & \multicolumn{2}{c|}{\ \ 30.64} & 6.19  & 0.16  & \multicolumn{2}{c|}{\ \ 29.33} \\
 \hline
 6.27-6.37 & 14.00  &       &       & \multicolumn{2}{c|}{} &       &       & \multicolumn{2}{c|}{} \\
 \hline
 \multirow{2}[0]{*}{6.40-6.85} & \multirow{2}[0]{*}{442.00} & \multirow{2}[0]{*}{6.62} & \multirow{2}[0]{*}{0.16} & \multicolumn{2}{c|}{\multirow{2}[0]{*}{444.94}} & 6.47  & 0.12  & 113.53  & \multirow{2}[0]{*}{395.06} \\
 \cline{7-9}
       &       &       &       & \multicolumn{2}{r|}{} & 6.66  & 0.13  & 281.53  &  \\
       \hline
 6.87-6.96 & 26.00  &       &       & \multicolumn{2}{r|}{} &       &       & \multicolumn{2}{r|}{} \\
 \hline
 \multirow{3}[0]{*}{6.98-7.81} & \multirow{3}[0]{*}{1425.00} & 7.19  & 0.16  & 528.45  & \multirow{3}[0]{*}{1335.57} & 7.09  & 0.13 & 254.19 & \multirow{3}[0]{*}{1233.05} \\
 \cline{3-5} \cline{7-9}
       &       & 7.49   & 0.16  & 721.79  &       & 7.33  & 0.15  & 496.30  &  \\
       \cline{3-5} \cline{7-9}
       &       &  7.71  &  0.11 & 85.33  &       & 7.59  & 0.16  & 482.56  &  \\
       \hline
\end{tabular}%
\end{table*}

\end{document}